\documentclass[useAMS,usenatbib]{mn2e}
\newif\ifPDF
\ifx\pdfoutput\undefined\PDFfalse
\else\ifnum\pdfoutput > 0 \PDFtrue
\else\PDFfalse
\fi
\fi

\ifPDF
\usepackage[pdftex]{graphicx}
\else
\usepackage[dvips]{graphicx}
\fi

\usepackage{amsmath}
\usepackage{amsfonts}
\usepackage{xspace}
\usepackage{url}

\newcommand{\eg}{e.g.\ }
\newcommand{\ie}{i.e.\ }
\newcommand{\hi}{H\,{\sc i}\xspace}
\newcommand{\hipass}{HIPASS\xspace}

\newcommand{\duchamp}{\protect\textsc{duchamp}\xspace}
\newcommand{\atrous}{\textit{{\`a} trous}\xspace}

\title[The \duchamp source finder]{{\Large\bf DUCHAMP}: a 3D source finder for spectral-line data}
\title[The {\normalsize DUCHAMP} source finder]{{\Large\bf DUCHAMP}: a 3D source finder for spectral-line data}

\author[Matthew Whiting]{Matthew T. Whiting $^1$\thanks{E-mail:
Matthew.Whiting@csiro.au} \\
$^1$ CSIRO Astronomy and Space Science, P.O. Box 76, Epping, NSW 1710, Australia}

\date{}

\begin{document}

\pagerange{\pageref{firstpage}--\pageref{lastpage}} \pubyear{2009}

\maketitle

\label{firstpage}

\begin{abstract}

This paper describes the \duchamp source finder, a piece of software
designed to find and describe sources in 3-dimensional,
spectral-line data cubes. \duchamp has been developed with \hi
(neutral hydrogen) observations in mind, but is widely applicable to
many types of astronomical images. It features efficient source
detection and handling methods, noise suppression via smoothing or
multi-resolution wavelet reconstruction, and a range of graphical
and text-based outputs to allow the user to understand the
detections. This paper details some of the key algorithms used, and
illustrates the effectiveness of the finder on different data sets.
\end{abstract}


\section{Introduction}

Large-scale blind surveys in astronomy provide a wealth of information
about the Universe. They are the best means to gain knowledge in an
unbiased way about populations of sources, particularly if the only
limitation is the flux limit imposed by observational constraints. For
determining the nature of the local Universe, neutral hydrogen (\hi)
surveys provide a view of the bulk of the baryonic mass, whilst also
giving insights into the dark matter content. The \hi Parkes All-Sky
Survey \citep[\hipass, ][]{barnes01-alt} is the best example of a truly
large-scale blind \hi survey, yielding catalogues of 4315 sources over
all the sky south of $\delta=+2^\circ$ \citep{meyer04-alt} and a
further 1002 sources between $+2^\circ<\delta<+25^\circ30\arcmin$
\citep{wong06-alt}, plus catalogues of the 1000 brightest galaxies
\citep{koribalski04-alt} and high-velocity clouds \citep{putman02-alt}. These
results provide a great deal of information on the \hi properties of
galaxies, the \hi mass function, the local velocity fields, and the
cosmic \hi mass density.

Blind surveys require a good method of detecting objects, to build up
a reliable catalogue that is as complete as possible. Such detection
techniques must cope with the presence of noise that provides spurious
sources and hides real but faint ones. The \hipass survey featured a
source detection procedure that required a large amount of human input
-- each source was verified several times by eye. As one increases the
size of the spectral cubes, in terms of spatial size and the number of
channels, this approach quickly becomes unfeasible, and a reliable
automated source finding algorithm becomes essential to extract the
required science catalogues from surveys.

An example of an instrument driving such development is the Australian
Square Kilometre Array Pathfinder \citep[ASKAP,][]{deboer09-alt}.
This will be a survey telescope, built in the radio-quiet SKA
candidate site of Western Australia, that will be able to do, amongst
other things, \hi surveys over large areas of sky and a wide redshift
range (due to its 30 square degree field of view and 300\,MHz
bandwidth). Ten ``Survey Science Projects'' (SSPs) have been
identified as the major projects to consume at least 75\% of ASKAP's
initial five years, and two of these are extragalactic \hi surveys:
the ``Widefield ASKAP L-band Legacy All-sky Blind surveY''
\citep[WALLABY,][]{koribalski09}, an all-sky survey to $z\sim0.26$
that is one of two top-ranked projects (along with the ``Evolutionary
Map of the Universe'' \citep[EMU,][]{norris11-alt} all-sky continuum
survey), and ``Deep Investigations into Neutral Gas Origins''
\citep[DINGO,][]{meyer09}, a deeper survey over a smaller area to
$z\sim0.5$. (Note that these are not the only SSPs with source finding
requirements, they are simply the two surveys for extragalactic \hi
emission.) The large data rates that will be produced by ASKAP will
necessitate largely automated processing, and so a robust, reliable
source finder will be essential if the surveys are to meet their
science goals.

The anticipated demands of ASKAP, combined with the realisation that
there was no publicly-available generic 3D source-finder suitable for
such work, prompted the development of the \duchamp source-finding
software package. This is a stand-alone piece of software, not closely
associated with any particular instrument or survey, but aims to be a
general-purpose 3D source finder. The algorithms, however, are being
evaluated by the ASKAP development team and the Survey Science teams
for inclusion in the processing pipelines for ASKAP.

\duchamp has been available for some years now, and has been used with
a variety of data from single-dish
\citep{wong07,mcdonnell08,putman09,hsu11,purcell12} and interferometer
\citep{johnston-hollitt08,lah09-alt,popping11} telescopes, as well as
an aid to visualisation techniques \citep{fluke10,hassan11}.  This
paper aims to demonstrate the utility of \duchamp by presenting some
of its key development aspects: the source detection techniques,
statistical calculations and noise-suppression techniques, as well as
memory management and software design. It also describes some of the
testing techniques that have been used in its evaluation. All examples
of \duchamp processing use version 1.1.13 of the software, the most
recent version at time of writing.

\section{The {\sevensize\bf DUCHAMP} source finder}
\label{sec-duchampintro}

\subsection{Philosophy and Design of {\sevensize\bf DUCHAMP}}
\label{sec-philosophy}

The philosophy I have used in developing \duchamp has been to make
things as friendly as possible for the user. This has resulted in
straightforward user input and a range of graphical and text-based
outputs to enable the user to accurately judge the quality of their
detections.

The aim of \duchamp is to provide the user with information about the
location of the sources of interest in the data. It does not make any
prior assumptions about the nature of those sources (\eg assuming they
are Gaussian-shaped), nor does it perform any profile fitting or
analysis beyond direct measurements of the provided data. The aim is
to provide the user with the \emph{location} of the interesting
sources. To that end, mask cubes can be written to highlight the
detected pixels, allowing the user to do their own post-processing.

\duchamp has been designed for use primarily with spectral-line cubes,
and this colours the thinking in much of the design. Spectral-line
cubes have two spatial dimensions and one spectral dimension
representing frequency, wavelength or velocity, and this distinction
is reflected in a number of areas, such as reference to ``channel
maps'', being 2D spatial slices at a particular spectral
value. Despite this, \duchamp is readily applicable to
two-dimensional, and even one-dimensional data.

The other main assumption made by \duchamp is that the sources are
sparsely populated throughout a cube that is dominated by noise. If
this assumption does not hold (for example, in the case of
observations of diffuse \hi from the Milky Way), processing will still
be able to be done, but any statistics \duchamp calculates will be
biased by the presence of signal. The noise calculations \duchamp
employs are detailed in Sec.~\ref{sec-threshold}.

\subsection{Obtaining and using the {\sevensize DUCHAMP} software}

\duchamp is open-source (GPL licensed) software, written in C++, and
is publicly available for download from the \duchamp web
site\footnote{http://www.atnf.csiro.au/people/Matthew.Whiting/Duchamp}. There,
one can find the source code, an extensive User's Guide and links to
required packages. These are:
\textsc{cfitsio}\footnote{http://heasarc.gsfc.nasa.gov/docs/software/fitsio/fitsio.html}\citep{pence99},
to handle operations on FITS files;
\textsc{wcslib}\footnote{http://www.atnf.csiro.au/people/mcalabre/WCS/index.html}\citep[described
in][]{calabretta02}, to correctly convert between pixel and world
coordinates; and
\textsc{pgplot}\footnote{http://www.astro.caltech.edu/$\sim$tjp/pgplot/},
to produce the graphical output. \duchamp should run in any
UNIX/Linux-based environment, and is being maintained to ensure this
continues to be the case. To this end, feedback is encouraged should
any problems arise when using it.

\subsection{Summary of processing}
\label{sec-summary}

\begin{figure}
\includegraphics[width=80mm]{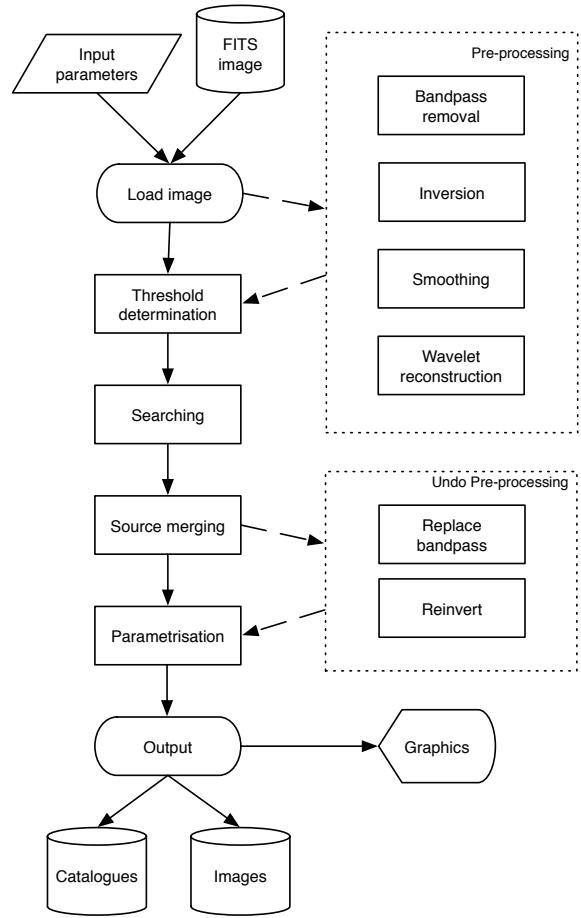}
\caption{A flow-chart indicating the key processing steps used by
\duchamp. The pre-processing options are user-selectable, and
are typically done before the determination of the threshold
used in searching.}
\label{fig-flow}
\end{figure}

This section summarises the different steps undertaken by \duchamp in
finding sources in an image. Certain elements of the processing will be
explained in detail in later sections. Fig.~\ref{fig-flow} indicates
the key processing steps that \duchamp will perform on an image.

\duchamp requires two inputs. The first is an image or cube, in FITS
format. The image merely needs to conform to the FITS standard
\citep{pence10} and have an IMAGE extension. FITS files produced by
common reduction packages, such as \textsc{miriad} and \textsc{casa},
should be suitably compatible. The second input is a set of
parameters that govern the execution. These parameters allow a large
degree of flexibility both in choosing different processing algorithms
and in fine tuning many aspects of the searching and output.

Following the reading of the image file, which can be the entire image
or a specified subsection (that is, the user can nominate pixel ranges
for each dimension of the image), a user-selectable set of
pre-processing can be done prior to searching. This can be simple
inversion of the image (to search for negative features), a basic
spectral bandpass subtraction to remove a continuum ripple or some
other large-scale spectral variation, or either smoothing or
multi-resolution wavelet reconstruction to enhance faint
sources. These latter two approaches are described in
Sec.~\ref{sec-noisesuppression}.

The threshold is then set according to the input parameters as either
a flux or signal-to-noise threshold - Sec.~\ref{sec-threshold}
provides details of how the threshold is determined. Sources are then
extracted from the image, and merged to form objects. These are then
parameterised - if relevant, the inversion and bandpass subtraction are
reversed prior to parametrisation. The detection, merging and
parametrisation steps are described in Sec.~\ref{sec-detection}.

The final object catalogue is then ready, and can be returned in a
variety of formats, including ASCII text and
VOTable/XML. Appendix~\ref{app-example} gives an example of the ASCII
output file. \duchamp also provides a range of graphical output to
assist the user in understanding the nature of the detected
sources. These include spatial maps showing the location and size of
each source, and individual plots for each source showing the both the
spectrum and the 0th-moment map.

Additional outputs include FITS images containing the mask indicating
the location of detected sources, or the smoothed or reconstructed
image (these FITS images will have their header information formatted
in the same way as the input image). These allow post-processing to
extract individual sources, or fit particular functions to detected
objects as dictated by the particular science of interest to the user.

\section{Representation of detected objects}
\label{sec-scan}

\begin{figure}
\includegraphics[width=80mm]{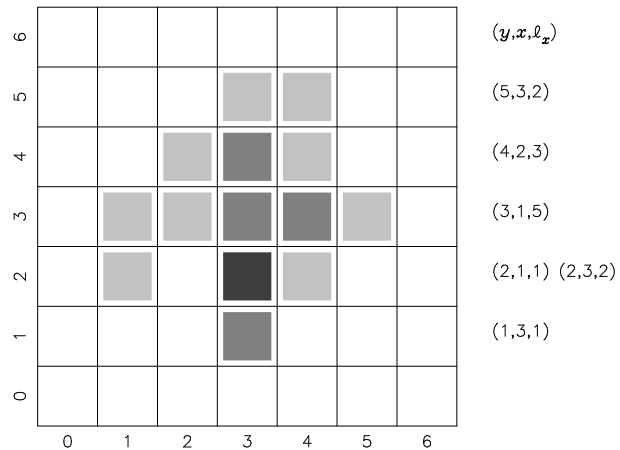}
\caption{An example of the run-length encoding method of storing
pixel information. The scans used to encode the image are listed
alongside the relevant row. The pixels are colour-coded by
nominal pixel values, but note that the pixel values themselves
do not form part of the encoding and are not kept as part of the
object class. }
\label{fig-objExample}
\end{figure}

To keep the memory usage manageable for the largest range of input
data possible, \duchamp implements specialised techniques for storing
the location of detected objects. A naive method could be to store
each single pixel, but this results in a lot of redundant information
being stored in memory.

To reduce the storage requirements, the run-length encoding method is
used for storing the spatial information. In this fashion, an object
in 2D is stored as a series of ``runs'', encoded by a row number (the
$y$-value), the starting column (the minimum $x$-value) and the run
length ($\ell_x$: the number of contiguous pixels in that row
connected to the starting pixel). A single set of $(y,x,\ell_x)$
values is called a ``scan''. A two-dimensional image is therefore made
up of a set of scans. An example can be seen in
Fig.~\ref{fig-objExample}. Note that the object shown has fourteen
pixels, and so would require 28 integers to record the positions of
all pixels. The run-length encoding uses just 18 integers to record
the same information. The longer the runs are in each scan, the
greater the saving of storage over the naive method.

A 3D object is stored as a set of channel maps, with a channel map
being a 2D plane with constant $z$-value. Each channel map is itself a
set of scans showing the $(x,y)$ position of the pixels.

\section{Statistical Analysis and Threshold Determination}
\label{sec-threshold}

\subsection{Thresholds in {\sevensize DUCHAMP}}

The fundamental step in source finding is the application of a
threshold, as this determines whether a given voxel\footnote{``Voxel''
refers to a ``volume element'' or ``volumetric pixel'' -- the 3D
equivalent of the 2D picture element or pixel.} is part of a source
(and therefore of some interest), or part of the background. \duchamp
uses a single threshold for the entire cube, leading to an output
catalogue with a uniform selection criterion. This approach lends
itself best to data that has uniform noise, but this need not be the
case: \duchamp has features that are able to remove some background
structure (such as large-scale spectral baseline features). The
ability to vary the threshold in response to the local noise would
allow deeper searching where the sensitivity warrants, and such
features are part of development under way for the ASKAP
source-finding pipeline.

The threshold for \duchamp can be given as a flux value directly, or
it can be expressed as a signal-to-noise level (a so-called
``$n\sigma$'' threshold). In this case, the value of $\sigma$ needs to
be measured from the image statistics. Either the entire set of image
pixels are used, or a subsection can be specified (via the parameter
file input) that specifies which pixels to measure the statistics
from. 

This latter option allows the user to, for instance, measure the noise
level in a region of the cube known to be free of sources (for
example, a frequency or velocity range in a spectral cube in which no
sources are expected). It may be, however, that no such region can be
found, and that the threshold has to be based on noise statistics
estimated from data containing sources (\ie voxels that are not pure
noise). \duchamp provides mechanisms to achieve this, either by using
robust methods to estimate the noise, or by setting the threshold in
an alternative way to the ``$n\sigma$'' approach. These methods are
detailed in the following sections.

\subsection{Robust techniques}

When searching for small objects in a noisy background detection
thresholds are often set at some multiple of the standard deviation of
the image background. However, direct measurements of this value can
be biased by the presence of bright pixels, yielding
overestimates. Furthermore, these bright pixels are often in the
sources of interest or some additional artefact (such as interference),
rather than the background, and so should not contribute to the
background calculation.

It can be preferable, then, to calculate the noise properties via
robust methods. \duchamp makes use of the median absolute deviation
from the median (henceforth MADFM) as a proxy for the standard
deviation. Although computationally expensive (it requires the data to
be at least partially sorted twice\footnote{Only partial sorting is
required, since the median requires just the middle element(s). An
algorithm such as \texttt{nth\_element} from the C++ Standard
Template Library is suitable.}), it is very robust against bright
pixels. It is best used in the situation where the data in question is
dominated by a large number of noise pixels, with a relatively small
number of brighter pixels present.  It should be noted that the
standard error of the median is 1.253 times that of the mean, and so
can be more susceptible to fluctuations of the sampling, but with
well-behaved noise this is acceptable.

Note that the assumption \duchamp makes about the nature of the data,
that there are relatively isolated sources embedded in data dominated
by noise, is important here. The median and MADFM will still be biased
by the presence of bright non-noise pixels, as what will be interpreted as
the centre of the distribution will be displaced from the centre of
the noise pixel distribution. The bias, however, only depends on the
number of non-noise pixels compared to the total size of the array
under consideration, and not on the actual values of the bright
pixels.

To compute the MADFM, $s$, one first calculates the median value of the array, and
then the median value of the absolute difference between the pixel
values and the median:
\begin{equation}
s = \rmn{med}\left(|x_i - \rmn{med}(x_i)|\right)
\end{equation}
For a pure Gaussian distribution of values,
the MADFM does not give the same value as the standard deviation, but
can be scaled easily to determine what the standard deviation of that
distribution is. This scaling can be calculated analytically. The
Normal, or Gaussian, distribution for mean $\mu$ and standard
deviation $\sigma$ has a density function
\[
f(x) = \frac{1}{\sqrt{2\pi\sigma^2}}\ e^{-(x-\mu)^2/2\sigma^2}.
\]
The median, $m$, is the middle of the distribution, defined for a
continuous distribution by
\[
\int_{-\infty}^{m} f(x) dx = \int_{m}^{\infty} f(x) dx.
\]
From symmetry (since $f(x-\mu)=f(\mu-x)$), we quickly see that for the
continuous Normal distribution, $m=\mu$. We consider the case
henceforth of $\mu=0$, without loss of generality.

To find the MADFM, $s$, we find the distribution of the absolute
deviation from the median, and then find the median of that
distribution. This distribution is given by
\begin{eqnarray*}
g(x) &= &{\text{distribution of }} |x|\\
&= &f(x) + f(-x),\ x\ge0\\
&= &\sqrt{\frac{2}{\pi\sigma^2}}\, e^{-x^2/2\sigma^2},\ x\ge0.
\end{eqnarray*}
So, the median absolute deviation from the median, $s$, is given by
\[
\int_{0}^{s} g(x) dx = \int_{s}^{\infty} g(x) dx.
\]
If we use the identity
\[
\int_{0}^{\infty}e^{-x^2/2\sigma^2} dx = \sqrt{\pi\sigma^2/2}
\]
we find that
\[
\int_{s}^{\infty} e^{-x^2/2\sigma^2} dx =
\sqrt{\pi\sigma^2/2}-\int_{0}^{s} e^{-x^2/2\sigma^2}dx.
\]
Hence,
\[
\int_{0}^{s}e^{-x^2/2\sigma^2} dx - \sqrt{\pi\sigma^2/8} = 0,
\]
which gives $s = \sqrt{2}\operatorname{erf}^{-1}(0.5)\sigma =
0.6744888\sigma$. Thus, to estimate $\sigma$ for a Normally distributed data
set, one can calculate $s$, then divide by 0.6744888 (or multiply by
1.4826042) to obtain the correct estimator.

This conversion can be compared to solutions quoted elsewhere. For
instance, \citet{meyer04-alt} use the MADFM to estimate the
noise in a cube, then quote a ``cube noise'' parameter defined by
$s\sqrt{\pi/2}$. Clearly this ``cube noise'' should not be interpreted
as the standard deviation of the noise (despite being called
Rms$_\text{cube}$ in the HICAT catalogue) - it is in fact only 82\% of
the standard deviation as calculated above.

These robust techniques can be used in \duchamp to provide accurate
estimates of the noise background. When thresholds are defined
according to a multiple of the standard deviation, the MADFM is always
converted to an equivalent standard deviation using the conversion
factor obtained above. Users are cautioned, however, that even these
robust techniques can be biased by non-noise signal. In the case that
the data being examined is purely noise plus some positive signal (or
artefacts), then the median will not fall in the middle of the noise
distribution, but slightly to the positive side, and similarly for the
MADFM. For typical data, however, where the real signals are sparsely
populated and outnumbered by noise voxels this bias will be small, and
certainly smaller than the bias arising from normal statistics (mean
and standard deviation).

It may also be the case that the noise is not precisely Gaussian. For
instance, \citet{popping11} consider WSRT interferometer images and
find an excess number of negative pixels over that expected from an
ideal Gaussian distribution. These are ascribed to interferometry
artefacts in the image, which are expected to be symmetric about zero
flux due to the purely relative nature of interferometric data when
auto-correlations are not included. In practice, if one knows, or can
estimate the scaling between MADFM and $\sigma$, then it is
straightforward to scale the requested threshold such that the
appropriate $n\sigma$ level is applied. These points highlight the
need for users to have some understanding of the nature of the data
being searched.

\subsection{False discovery rate technique}

Although one can estimate the standard deviation more accurately with
the robust techniques, the problem of false discoveries will still be
present. A technique has been presented \citep{miller01,hopkins02}
that attempts to control the false discovery rate (FDR) in the
presence of noise. This fixes the false discovery rate (given by the
number of false detections divided by the total number of detections)
to a certain value $\alpha$ (\eg $\alpha=0.05$ implies 5\% of
detections are false positives). In practice, an $\alpha$ value is
chosen, and the ensemble average FDR (\ie $\langle FDR \rangle$) when
the method is used will be less than $\alpha$.  One calculates $p$ --
the probability, assuming the null hypothesis is true, of obtaining a
test statistic as extreme as the pixel value (the observed test
statistic) -- for each pixel (in \duchamp, Normal statistics are
assumed), and sorts them in increasing order. One then calculates $d$
where
\[
d = \max_j \left\{ j : P_j < \frac{j\alpha}{c_N N} \right\},
\]
and then rejects all hypotheses whose $p$-values are less than or
equal to $P_d$. (So a $P_i<P_d$ will be rejected even if $P_i \geq
j\alpha/c_N N$.) Note that ``reject hypothesis'' here means ``accept
the pixel as an object pixel'' (\ie we are rejecting the null
hypothesis that the pixel belongs to the background).

The $c_N$ value here is a normalisation constant that depends on the
correlated nature of the pixel values. If all the pixels are
uncorrelated, then $c_N=1$. If $N$ pixels are correlated, then their
tests will be dependent on each other, and so $c_N = \sum_{i=1}^N
i^{-1}$. \citet{hopkins02} consider real radio data, where the pixels
are correlated over the beam. For the case of three-dimensional
spectral data, the beam size would be multiplied by the number of
correlated channels. This number can be specified via the input
parameter set\footnote{Using the parameter \texttt{FDRnumCorChan}.},
and so a user can set it according to their knowledge of the cube
being searched. This can be important if the cube is known to have
been smoothed prior to searching, or if the telescope backend results
in some correlation between neighbouring channels.

The theory behind the FDR method implies a direct connection between
the choice of $\alpha$ and the fraction of detections that will be
false positives. These detections, however, are individual pixels,
which undergo a process of merging and rejection, and so the fraction
of the final list of detected objects that are false positives will
often be much smaller than $\alpha$.

\section{Detection of Objects}
\label{sec-detection}

\subsection{Searching strategy}

The problem of identifying sources in a two-dimensional image, by
means of a single threshold, has well defined solutions. \citet{lutz80}
presented an algorithm for locating objects in an image with a single
raster-scan pass through the image. Such algorithms are possible
because of the well-nested nature of objects in two dimensions: on a
given row of an image, if object A lies between two portions of object
B, then \textit{all} of object A lies between those two portions.

The problem of 3-dimensional source extraction is, in general, a more
complex one. The extra dimension means that the well-nested property
no longer applies, and so distinct objects can be intertwined while
still remaining distinct. This makes it hard for a simple
raster-scanning (or equivalent) algorithm to be applied.

The approach \duchamp uses has two aspects: multiple lower-dimensional
searches, followed by merging of the detected sources to produce
complete three-dimensional sources. The simplest search \duchamp can
do is to treat each spatial pixel as a distinct spectrum, search with
a simple 1D search routine, and combine the objects afterwards. The
alternative (actually the default) is to treat each frequency channel
as a two-dimensional image, and search it using the algorithm of
\citet{lutz80}. The algorithm performs a raster scan of the image,
processing one horizontal line at a time and builds up a list of
objects by keeping track of which objects in each row are connected
(in an 8-fold manner) to objects on the previous row. Objects from
different images are then combined afterwards (see
Sec.~\ref{sec-merge}).

\duchamp has been developed for three-dimensional data, and even
though we are using two- or one-dimensional searching, we know that
the objects we are interested in will be three dimensional (that is,
extended in all three directions). Ideally one would want to make use
of this fact to help identify sources. A good way to do this is to use
the smoothing or wavelet reconstruction approaches introduced
in Sec.~\ref{sec-noisesuppression}. Generally, smoothing in a direction perpendicular to the
direction of the searching will respond well to 3D structures. For
instance, if a 3D source is smoothed spectrally, then its brightness
in each 2D channel map will be enhanced relative to the background due
to the flux in neighbouring channels. Searching in 2D will then be
more sensitive than a search without the prior smoothing.

A single detection threshold is used at this point, effectively
treating each voxel in a binary manner. Once a source list has been
established (see Sec.~\ref{sec-merge}), the objects can be ``grown''
to a secondary threshold if required. This enables the detection of
all faint features provided they have at least one (or some number --
the minimum size of a detection is a user-selectable parameter) voxel
above the primary threshold.

\subsection{Merging of sources}
\label{sec-merge}

To form the final catalogue, objects found in individual channel maps
or spectra need to be merged. Merging takes place if a pair of objects
are judged to be close according to criteria determined by the
user. Either the objects must be adjacent, meaning there is at least
one pair of voxels (one voxel from each object) that are adjacent in
one direction, or they must have a separation less than a given number
of voxels (which can differ between spatial and spectral
separations).

The merging is done in two stages. When a single search of a 2D plane
or a 1D spectrum is complete, each detected object (stored as a set of
scans) is added to a master list of detections. It is first compared
to all those present in the list, and if it is close to one (according
to the criteria defined by the user) the two objects are merged. No
other comparisons are made at this point (to save time). If it is
close to none, it is added to the end of the list.

Once all the searching has been completed, the list needs to be
examined for any further pairs of close objects. Although an initial
comparison has been done, there will be unexamined pairs present. For
instance, consider objects A and B, which are not close, and a new
object C that is close to both A \& B. When C is added to the list, it
is first compared to A, found to be close, and combined. The list then
has the combined AC object and B, which are both close. A second
comparison needs to be made to combine these two. This secondary
merging stage looks at successive pairs of objects from the list. If
they are close, the second object is added to the first and removed
from the list. This ensures that all possible combinations are found.

The way the comparison is done is to first look at the pixel ranges
(minimum \& maximum in each dimension) to see if they overlap or lie
suitably close according to the above criteria. This is a quick way of
ignoring objects that should not be merged without doing a detailed
examination.  If they do, then the individual scans (refer
Sec.~\ref{sec-scan}) from common channels in each object are examined
until a pair is found that satisfy the criteria for closeness. Note
that this can be done directly on scans, without needing to look at
individual pixels.

If growing of sources is requested, this is done after the merging
step. A second merge is then performed to ensure that no duplicate
detections are present, due to sources growing over each other.

\subsection{Parametrisation of sources}

Once the unique set of sources has been found, \duchamp measures a
basic set of parameters for them. The location of the source is
measured in three ways: the peak position, or the voxel containing the
maximum flux value; the ``average'' position, being the average of all
pixel values in each dimension
\begin{equation}
\label{eq:1}
(\bar{x},\bar{y},\bar{z})=\left(\frac{\sum x_i}{N},\frac{\sum y_i}{N},\frac{\sum z_i}{N}\right)
\end{equation}
where $N$ is the number of pixels; and
the ``centroid'' position, being the flux-weighted average of all
pixels in each dimension
\begin{equation}
\label{eq:3}
(x_c,y_c,z_c)=\left(\frac{\sum(F_i x_i)}{F_T},\frac{\sum(F_i y_i)}{F_T},\frac{\sum(F_i z_i)}{F_T}\right)
\end{equation}
where $F_T=\sum F_i$ is the total flux of all detected pixels.

All positions and spectral locations are given in both pixel and world
coordinates, making use of the
\textsc{wcslib}\footnote{http://www.atnf.csiro.au/people/mcalabre/WCS/index.html}
library \citep[described in][]{calabretta02}. The sizes of the sources
are reported: in the spatial direction, this is just the extent of the
detected source (although not every pixel within this extent is
necessarily detected -- if the closeness criteria permit it, there can
be voxels below the threshold within this range).

In the spectral direction the parameters W50 and W20 are provided,
being the width at 50\% and 20\% of the peak flux. These are measured
on the integrated spectrum (\ie the spectra of all detected spatial
pixels summed together), and are defined by starting at the outer
spectral extent of the object (the highest and lowest
spectral values) and moving in or out until the required
flux threshold is reached. The widths are then just the difference
between the two values obtained. The full spectral width of the
detection (WVEL) is also reported -- this is analogous to the spatial
extent, being the extent of the detected source in the spectral
direction.

The flux of the source is given by the peak flux and both the ``total
flux'', defined above as $F_T$, and the ``integrated flux'' $F_I$, or
the total flux integrated over the spectral extent and corrected for
the beam:
\[
F_I = \frac{\sum F_i \Delta v_i}{B}
\]
where $B=\pi \alpha \beta / 4 \ln(2)$ is the area of a beam of major
and minor axes $\alpha$ and $\beta$ (in pixels), and $\Delta v_i$ is
the spectral width of each voxel.

These parameters are intended as a basic set, as \duchamp has not been
optimised for any particular survey. Those wanting more detailed or
science-specific parameters should make use of the mask cubes (as
discussed in Sec.~\ref{sec-philosophy}) to extract the locations of
detected objects and do their own processing from there.

Note that these parameters are measured using just the detected
pixels, since \duchamp does no fitting of spectral or spatial
profiles. This means that some parameters, integrated flux for
instance, will be biased by not including voxels that fall below the
detection (or secondary) threshold.  The exceptions to this are W50
and W20, as these are calculated using the full integrated spectrum,
including channels not necessarily forming part of the detection.

\section{Noise Suppression}
\label{sec-noisesuppression}

The presence of noise ultimately limits the ability to detect
objects. This occurs in two ways: reducing the completeness of the
final catalogue by making faint sources appear to have a flux
erroneously below the detection threshold; and reducing the catalogue
reliability by generating false detections through detection of
spurious noise peaks. \duchamp implements several methods to minimise
the effect of the noise and improve the reliability, which are
discussed here.

\subsection{Smoothing}
\label{sec-smooth}

\begin{figure*}
\begin{minipage}{170mm}
\includegraphics{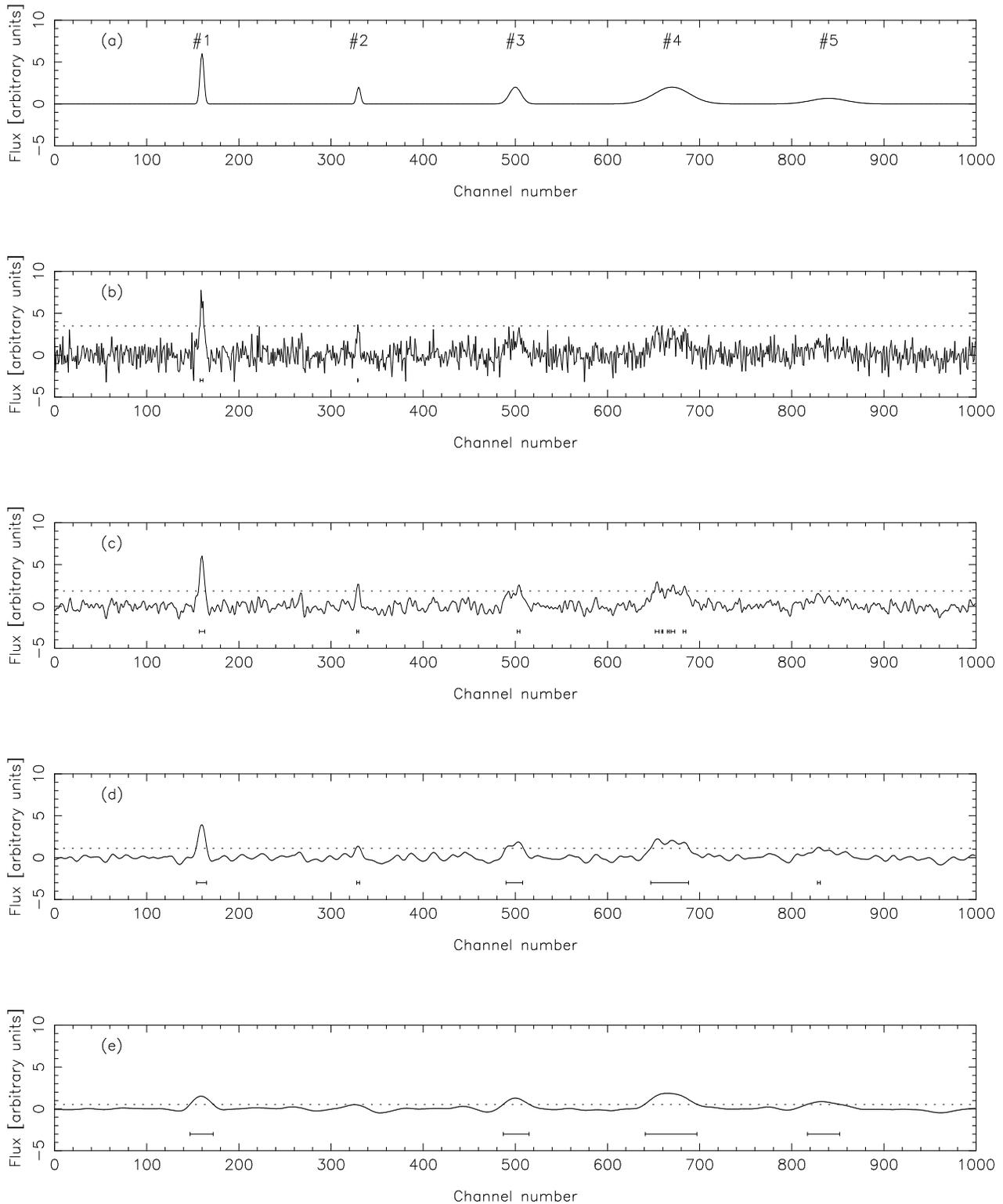}
\caption{An illustration of the effect of smoothing on the
detectability of sources. See text for description. (a) A
1-dimensional model spectrum, with five Gaussian sources (labels
are referred to in the text). The central three have the same
peak but different FWHM: 5, 15 and 45 channels. The first and
last have the same integrated flux as the central peak. (b) The
model spectrum with random noise added, drawn from a standard
normal distribution. A $3\sigma$ threshold is indicated by the
dotted line, and the bars under the spectrum indicate contiguous
sources above the threshold. (c) The noisy spectrum smoothed
with a width of 5 pixels, with the $3\sigma$ threshold, as
measured from the spectrum, indicated. (d) As for (c), but
smoothed with a width of 15 channels. (e) As for (c), but
smoothed with a width of 45 channels.}
\label{fig-smoothExample}
\end{minipage}
\end{figure*}

One option is to smooth the data, either spectrally (via a Hanning
filter), or spatially (via convolution with a Gaussian kernel). When
the size of the filter/kernel is chosen appropriately, these
approaches can be very effective at enhancing structures of a
particular scale, and suppressing the finer-scale noise
fluctuations. They are limited, however, to enhancing just the scale
determined by the filter size. This becomes an \textit{a priori}
choice made by the user, which may be appropriate if searching for a
particular type of signal (for instance, galaxies of a particular
velocity width), but can bias the results of a truly blind search by
suppressing other scales.

To examine how the choice of scale influences the outcome,
Fig.~\ref{fig-smoothExample} shows the effect of smoothing on
differently-sized sources (just in one dimension, for ease of
representation). The model spectrum, shown in (a) and in (b) (the
latter with noise added), contains five sources. The flux scale is
chosen so that the noise has a standard deviation of 1 unit. The three
central sources \#2, \#3 \& \#4 each have the same peak flux of 2 with
different Full Width at Half Maximum (FWHM) values: 5, 15 and 45
channels respectively, while \#1 and \#5 have the same integrated flux
as \#3, but the same width as \#2 and \#4 respectively.  Panel (b)
shows this spectrum with noise added, and with a $3\sigma$ threshold
superimposed. The value of $\sigma$ is that measured from the spectrum
itself, and so is a little larger than 1 due to the presence of
positive sources(in this example it is 1.11). Below the spectrum we
indicate with bars where we detect sources above this threshold. Only
the first two sources are detected -- the second perhaps due to a
boost from the noise.

Panel (c) shows the spectrum after smoothing with a Hanning filter of
the same width as the first two sources. This keeps the first two
sources visible, and makes enough structure visible in sources \#3 and
\#4 for them to be detected, although note that \#4 is
fragmented. When smoothing with the width of source \#2 (panel (d)),
all five sources are able to be recovered, with source \#4 now forming
a single detection. By the time we smooth with the width of sources
\#4 and \#5 (panel (e)), the weakness of \#2 results in it not being
detected - we are starting to smooth over structure on its scale (\#1
is still detected, but its peak strength is greatly reduced).

\subsection{Wavelet reconstruction}

\subsubsection{Algorithm}
\label{sec-atrous}

\begin{figure*}
\begin{minipage}{170mm}
\includegraphics{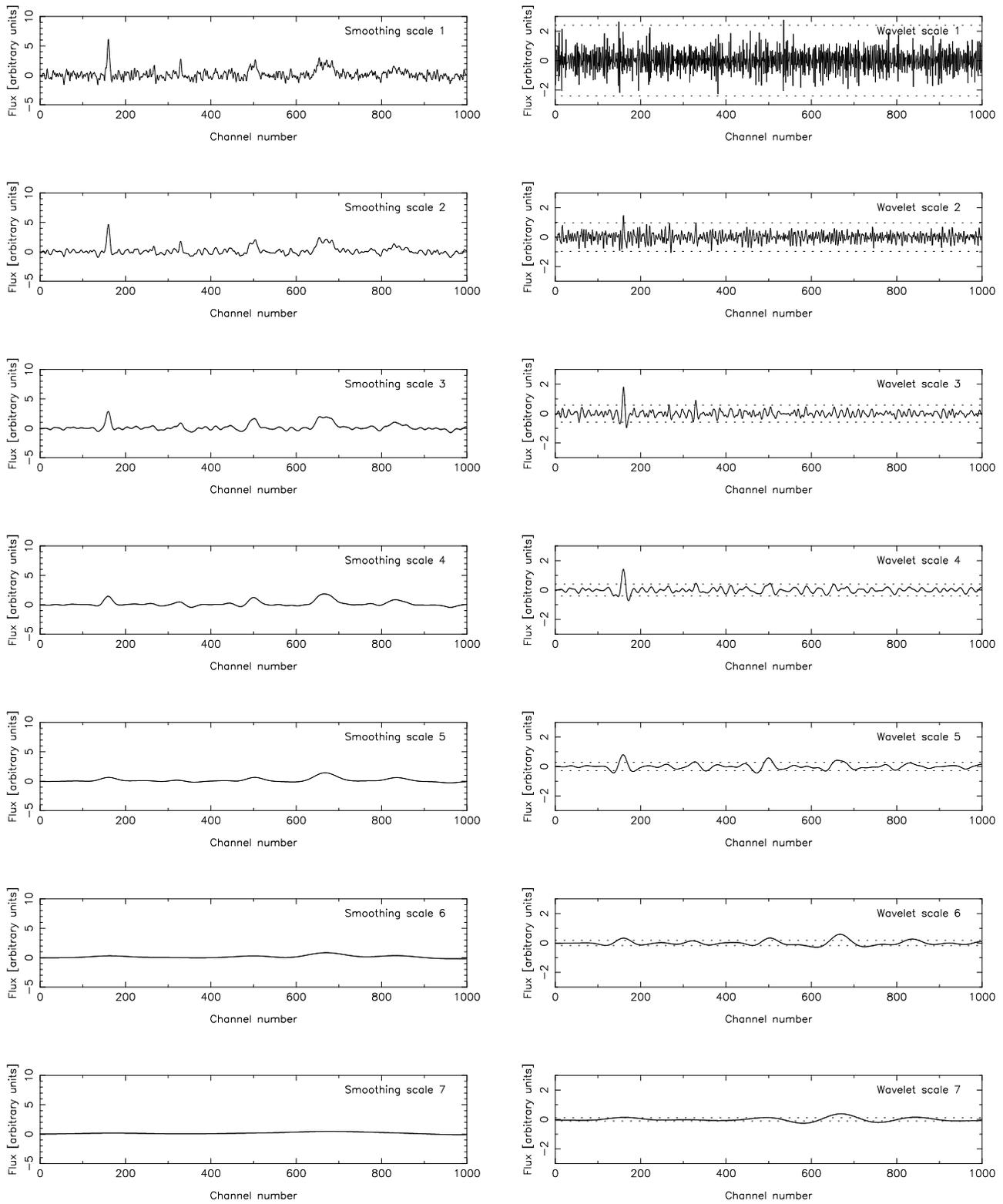}
\caption{A detailed view of what happens at each scale during the
reconstruction. The left-hand column shows the spectrum smoothed
at successively larger scales, all with the same vertical axis
range. The right-hand column shows the wavelet spectra, being
the difference between the smoothed spectrum to the left and the
one preceding it (or, in the case of the first scale, the first
smoothed spectrum and the original spectrum). Also shown on the
wavelet spectra, as dotted lines, are the $3\sigma$ thresholds
used in the reconstruction. }
\label{fig-reconExampleScales}
\end{minipage}
\end{figure*}

Clearly, smoothing with a given filter imposes a choice of scales for
which the search is sensitive, but this may result in information on
other scales (either finer or broader) being lost. An alternative
approach is to use a multi-resolution wavelet transform that works
over a range of scales. \duchamp uses the \atrous, or ``with holes'',
transform \citep{starck94} to sample the structure on scales at
logarithmically-spaced intervals. The algorithm used in the \atrous
transform is as follows:
\begin{enumerate}
\renewcommand{\theenumi}{(\arabic{enumi})}
\item Initialise the reconstructed array to zero everywhere.
\item Discretely convolve the input array with the chosen filter
function.
\item Define the wavelet coefficients by taking the difference between
the convolved array and the input array (input $-$ convolved).
\item Threshold the wavelet coefficients array: only keep those values
above a designated threshold, and add them to the reconstructed
array.
\item Double the separation between elements of the filter function,
creating the holes in the filter coverage alluded to by the
transform name.
\item Repeat from step 2, using the convolved array instead of the
input array.
\item Continue until the required maximum number of scales is
reached. This is defined by the largest scale where the filter
function size is not greater than the array size.
\item Add the final smoothed (\ie convolved) array to the
reconstructed array. This provides the ``DC offset'', as each of the
wavelet coefficient arrays will have zero mean.
\end{enumerate}

The doubling of the filter scale means that a range of scales from the
finest to the broadest are sampled in creating the reconstructed
array. The \atrous transform is a so-called ``redundant''
transform. This means that the sum of all wavelet arrays, plus the
final smoothed array, results in the input array. Symbolically,
denoting the input and output arrays by $I$ and $O$, and the smoothed
and wavelet arrays at scale $i$ by $S_i$ and $W_i$, this can be written
\begin{eqnarray*}
\label{eq:2}
O &= &\sum W_i + S_n \nonumber\\
&= &(I-S_1) + (S_1-S_2) + ... + (S_{n-1}-S_n) + S_n \nonumber\\
&= &I
\end{eqnarray*}

Applying a threshold
to the wavelet coefficients has the effect of only keeping those parts
of the input array where there is significant signal at some scale. If
the threshold is chosen well, the random noise present in astronomical
images can be suppressed quite effectively.
The threshold applied is a constant multiple (provided by the user) of
the noise level in the wavelet array.

The reconstruction may be done in one, two or three dimensions. The
three-dimensional case reconstructs the full array at once, by
defining a filter function that has the same profile in each of the
three dimensions. A single largest scale is used, limited by the
smallest array dimension. In the one-dimensional case, the 1D spectrum
for each spatial pixel is reconstructed independently, while the
two-dimensional case sees each 2D channel map reconstructed
independently. Examples of reconstructions with different
dimensionality are discussed in Sec.~\ref{sec-reconExample} below.

It was shown in Sec.~\ref{sec-smooth} and Fig.~\ref{fig-smoothExample}
how smoothing reduces the measured noise in a spectrum, by making
pixels more and more correlated with their neighbours. The implication
for the wavelet reconstruction of this is to reduce the measured noise
level at each scale, affecting how the threshold is applied. Since we
need to know each scale's noise, to apply a fixed signal-to-noise
threshold, we need to calculate it for each wavelet array. Rather than
doing so directly, we scale the noise from that measured in the input
array to take into account the increasing correlation of the pixels.

To do this, for a given filter function, we transform a spectrum of
zero flux save for a single unit pixel in the centre. By summing in
quadrature the pixel values at each wavelet scale, we can find the
scaling factor for the standard deviation. This procedure is explained
in detail in Appendix~\ref{app-scaling}.

These calculations depend on the pixels in the original array being
independent. If they are not (for instance, the beam or point spread
function of a telescope will result in partially correlated
neighbouring pixels, or neighbouring channels may be partly correlated
by the instrument or by some processing prior to \duchamp analysis),
taking the sum in quadrature is not the appropriate course of
action. In this case, the only way to obtain the correct scaling
factors would be through simulations that properly account for the
correlations in the pixels.

\subsubsection{Example}
\label{sec-reconExample}

\begin{figure*}
\begin{minipage}{170mm}
\includegraphics{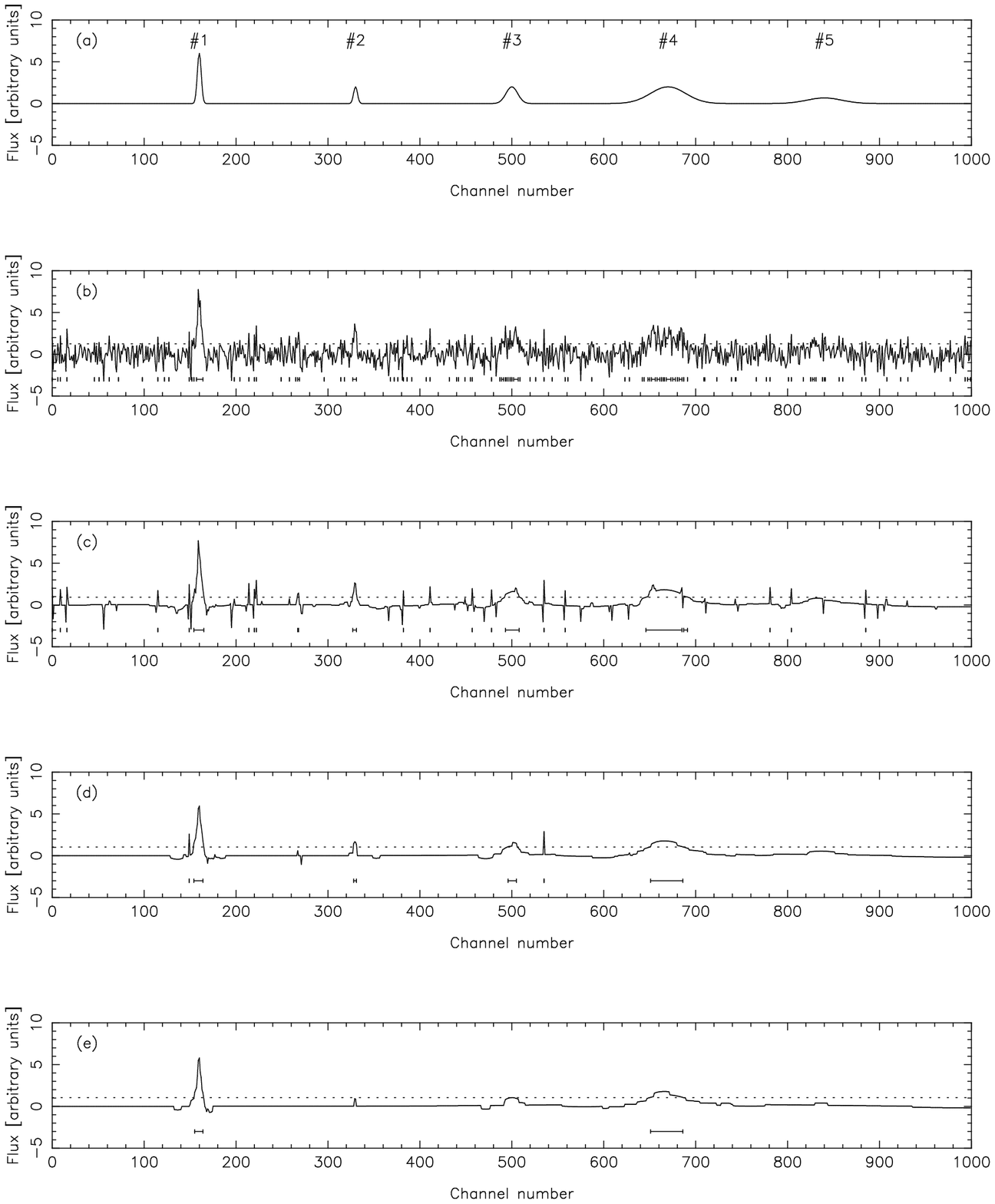}
\caption{An illustration of the wavelet reconstruction, using the
same data as for Fig.~\ref{fig-smoothExample}, with three
different values of the threshold applied to the wavelet
components. (a) The same model spectrum as used in
Fig.~\ref{fig-smoothExample}a), with the sources marked. (b) The
input spectrum (again identical to Fig.~\ref{fig-smoothExample})
with a $1\sigma$ detection threshold indicated by the dotted
line. The resulting detected sources are shown under the
spectrum. The remaining three panels show wavelet reconstruction
using a reconstruction threshold of: (c) $2\sigma$ (d) $3\sigma$
(the results of the reconstruction shown in detail in
Fig.~\ref{fig-reconExampleScales}) and (e) $4\sigma$.}
\label{fig-reconExampleResults}
\end{minipage}
\end{figure*}

\begin{figure*}
\begin{minipage}{170mm}
\includegraphics{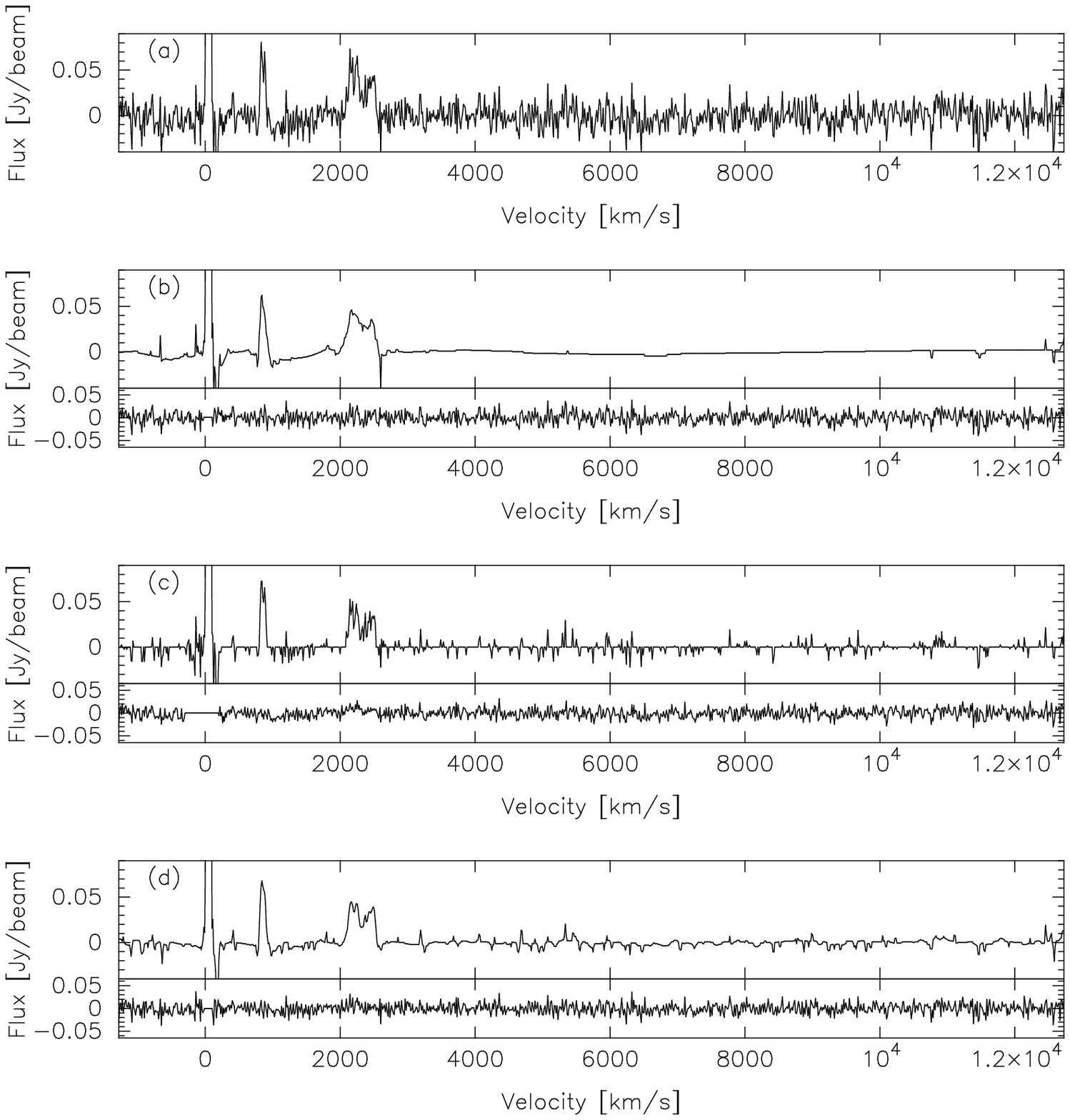}
\caption{A comparison of wavelet reconstruction with different
dimensionality. (a) The input spectrum (b) 1D reconstruction (c)
2D reconstruction (d) 3D reconstruction. For the reconstructed
spectra, the bottom graph shows the residual (input $-$
reconstructed).  The same wavelet threshold, $3\sigma$, was used
in each case. The input data is a cube from \hipass (cube \#201,
at $\rmn{RA}=6^{\rmn{h}}~12^{\rmn{m}}~2\fs9$,
$\rmn{Dec.}=-21^\circ~42'~24\farcs 8$) that shows two galaxies
in the one spectrum. Only a single 1D spectrum is shown (from a
single spatial pixel), but the full cube was used in the
reconstruction. The vertical axis limits have been chosen to
highlight the majority of the spectrum, but this means the
bright Galactic HI at channel~100 is truncated.}
\label{fig-reconExampleDim}
\end{minipage}
\end{figure*}

In Fig.~\ref{fig-reconExampleScales}, the step-by-step detail of a
reconstruction is shown, highlighting what happens to the smoothed and
wavelet spectra at each scale. The left hand side of the figure shows
the spectrum smoothed at successively larger scales -- the effects
described in Sec.~\ref{sec-smooth} are apparent here too. An 8th
smoothed spectrum, the ``DC offset'' described above, is also
computed, although, for simplicity, is not plotted (it is a fairly flat
line and has no matching right hand side column).

The right hand side of Fig.~\ref{fig-reconExampleScales} shows how the
wavelet spectrum changes from scale to scale. These spectra show where
the strongest signal (either positive or negative) can be found for a
given scale - note how source \#1 is strongest between scales 3 and 5,
\#3 strongest at scales 5 and 6, and \#4 at 6 and 7. Also shown on the
wavelet spectra are the thresholds used in the reconstruction. These
are set at $\pm3\sigma$, where $\sigma$ for each spectrum is scaled in
the manner described above from that measured in the input
spectrum. Only data outside the dotted lines are added together (along
with the DC offset) to create the reconstructed spectrum.

The effectiveness of the wavelet reconstruction are evident in,
Fig.~\ref{fig-reconExampleResults}, where the results of
reconstructing the model spectrum used in Fig.~\ref{fig-smoothExample}
are shown. Three wavelet threshold levels are used: 2, 3 (as used in
Fig.~\ref{fig-reconExampleScales}) and 4 $\sigma$. Clearly, as the
threshold is increased, more of the fine-scale noise is removed. Note,
however, that the sources, particularly the fainter, narrower ones,
get affected as well, so that the $4\sigma$ reconstruction starts to
lose sources \#2 and \#3.

In setting the threshold, the wavelet reconstruction requires a
different approach to the smoothing case. A signal to noise threshold
is determined based on the statistics measured from the residual
spectrum. If the wavelet reconstruction worked perfectly, this would
contain only the noise, and so would give the exact measurement of the
image noise. But the threshold is applied to the reconstructed
spectrum, which, ideally, would hold just the sources of
interest. Since there is (ideally) no noise present, the threshold can
be set much lower than one would otherwise consider. The example in
Fig.~\ref{fig-reconExampleResults} shows a $1\sigma$ threshold applied
to all cases. When applied to the raw spectrum, a vast number of
clearly spurious sources is returned, but the reconstructions remove
the bulk of these spurious sources.  This can be taken to extremes --
a $0.2\sigma$ detection threshold even allows recovery of source \#5
for all cases, with only 4 and 1 additional (spurious) sources for the
$3\sigma$ and $4\sigma$ reconstructions respectively.

Since the residuals have had structures that are not noise removed,
estimating the noise properties from them provide much better
estimates. The values for the $1\sigma$ threshold in each of the
spectra in Fig.~\ref{fig-reconExampleResults} are 1.25 for the input
spectrum, compared to 0.92, 1.03 and 1.06 for the $2\sigma$, $3\sigma$
and $4\sigma$ results respectively (these values include the mean --
the robust estimates for the standard deviation of the noise are 1.11
for the input, and 0.89, 0.99 and 1.03 for the respective
reconstructions).

In Fig.~\ref{fig-reconExampleDim}, we show the effect of changing the
dimensionality of the reconstruction. For this, we use real data from
the \hipass survey \citep{barnes01-alt}. Shown is a single spectrum,
taken from the \hipass cube \#201 at
$\rmn{RA}=6^{\rmn{h}}~12^{\rmn{m}}~2\fs9$,
$\rmn{Dec.}=-21^\circ~42'~24\farcs 8$. This position is chosen to
highlight two galaxies at velocities of 861 km~s$^{-1}$ (UGCA 120) and 2319
km~s$^{-1}$ (NGC 2196) which have quite different spectral profiles - this
highlights how the reconstruction recovers structure on different
scales. Three reconstructions are performed, using the one-, two- and
three-dimensional algorithms. Clearly, all three approaches
recover both galaxies, while removing a large proportion of the noise.

There are differences, however. There is a larger degree of
channel-to-channel noise remaining in the 2D reconstruction, and it
appears that more of the input spectrum is included in the
reconstructed spectrum (see around $v\sim0$km~s$^{-1}$ and in the range of the
broader galaxy). This appears to be largely due to the way the
reconstruction has been done. The 2D reconstruction works on each
channel map separately, so each channel in the extracted spectrum has
been calculated independently, depending not on its neighbouring
channels but the structure in the channel map. The channel maps are
also where the largest pixel-to-pixel correlations exist (that is,
within a beam), which allows beam-sized noise fluctuations to rise
above the reconstruction threshold more frequently than in the 1D
case.

The 3D case, while still containing more noise than the 1D (for the
same reason - it is also seeing the spatially-correlated beam noise),
does seem to give a better-looking reconstruction of the shape of both
the galaxies and the spectral baseline (the sharp negative features
next to the galaxies in the 1D spectrum are less well-defined in the
3D spectrum). The 3D reconstruction uses the full range of information
for a given pixel, so that structure in all three directions
contributes to the information about its reconstructed value.

\section{Evaluating {\sevensize\bf DUCHAMP}}

\subsection{Testing}

The aim of this section is to provide an indication of the relative
performance of the different algorithms of \duchamp. This is done
through simple tests using synthetic data purely to provide indicative
performance. Synthetic cubes were used for this purpose, rather than
real survey data from, say, \hipass, as this prevents systematic
effects that limit the completeness of the survey from affecting the
interpretation of the completeness of the source-finder.  A complete
set of tests that is relevant for a particular survey would use
simulations and/or real data that better match the noise and source
characteristics expected for that survey, and so is beyond the scope
of this paper.

The synthetic data used for these tests were cubes of dimensions
$128\times128\times512$ voxels, with a grid of sources of varying
brightnesses. We use 16 sources, ranging in peak brightness from 1.5
to 9 units, increasing in increments of 0.5. Each source is spatially
unresolved, so that it lies in a single pixel, and has a spectral
shape of a Gaussian with a 5 channel FWHM. The locations of the
sources are chosen so that there is at most one source in a given
channel map or spectrum.

The cube then has noise added to each voxel, sampled from a normal
distribution with a standard deviation of 1 flux unit (meaning the
flux units in the cube are essentially units of $\sigma$), and is then
convolved with a Gaussian beam that has a 3-pixel FWHM. This cube is
then searched for sources. This process is repeated 100 times for
different realisations of the noise, to provide an idea of the average
completeness and reliability of the algorithms.

We also make a reference cube without noise, but with the same
beam. This provides the ``truth'' result - what would be found in the
ideal case of no noise. This reference cube is searched both to a
depth of 0.01 (providing the full extent of sources), and to the same
depth as searches in the noisy cubes.

The cubes produced in this manner are then able to be searched using
different \duchamp parameter sets. Each parameter set will correspond
to a different mode of operation: basic searching, smoothing or
wavelet reconstruction, with different requirements on the minimum
number of spatial pixels \& spectral channels, as well as different
smoothing widths or reconstruction thresholds. Each parameter set is
then applied with a range of input signal-to-noise thresholds: 3, 4,
5, 6, 7 and 8. A given signal-to-noise threshold will provide
different effective flux thresholds when different pre-processing is done:
Note that the different pre-processing will result in
different effective flux thresholds for a given signal-to-noise: in
the smoothing case, the noise is measured from the smoothed cube,
where it is lower due to the correlated pixels, and the threshold is
applied to the same cube; while in the reconstruction case the noise
is measured from the rejected pixels (which ideally will indeed be
noise), and the threshold is applied to the reconstructed pixels.

Searching each of the 100 cubes, and then cross-matching with the
truth result, allows us to find the average number of times each
source is detected, subject to the noise. In this way, a completeness
curve can be constructed for a given search threshold, showing the
fraction of sources detected as a function of peak source
brightness. We can also determine the reliability for a given search
threshold, being the fraction of detected sources that correspond to a
true source.

The reliability measure here is implicitly dependent on the image
size. In the case of these simulations, a larger image size would
result in more noise detections with no additional true sources,
leading to a decrease in reliability. The key aspect to look at when
examining the results of the tests is the different dependence of the
reliability on the detection threshold from test to test.

\subsection{Evaluation of Results}

\begin{figure}
\includegraphics{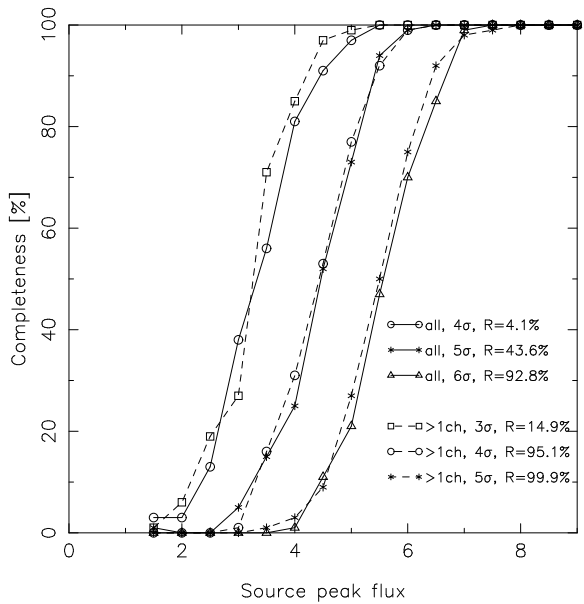}
\caption{Comparison of the completeness as a function of the peak
flux of sources for different detection thresholds and different
rejection techniques. The solid lines show searches at (from
left-to-right) $4\sigma$, $5\sigma$ and $6\sigma$, with no
restriction on the size of objects. The dashed lines show searches
at $3\sigma$, $4\sigma$ and $5\sigma$, rejecting sources that span
only one channel. The reliability ($R$, the fraction of detected
sources that are real) are indicated for each search type.}
\label{fig-comp-basic}
\end{figure}

\begin{figure}
\includegraphics{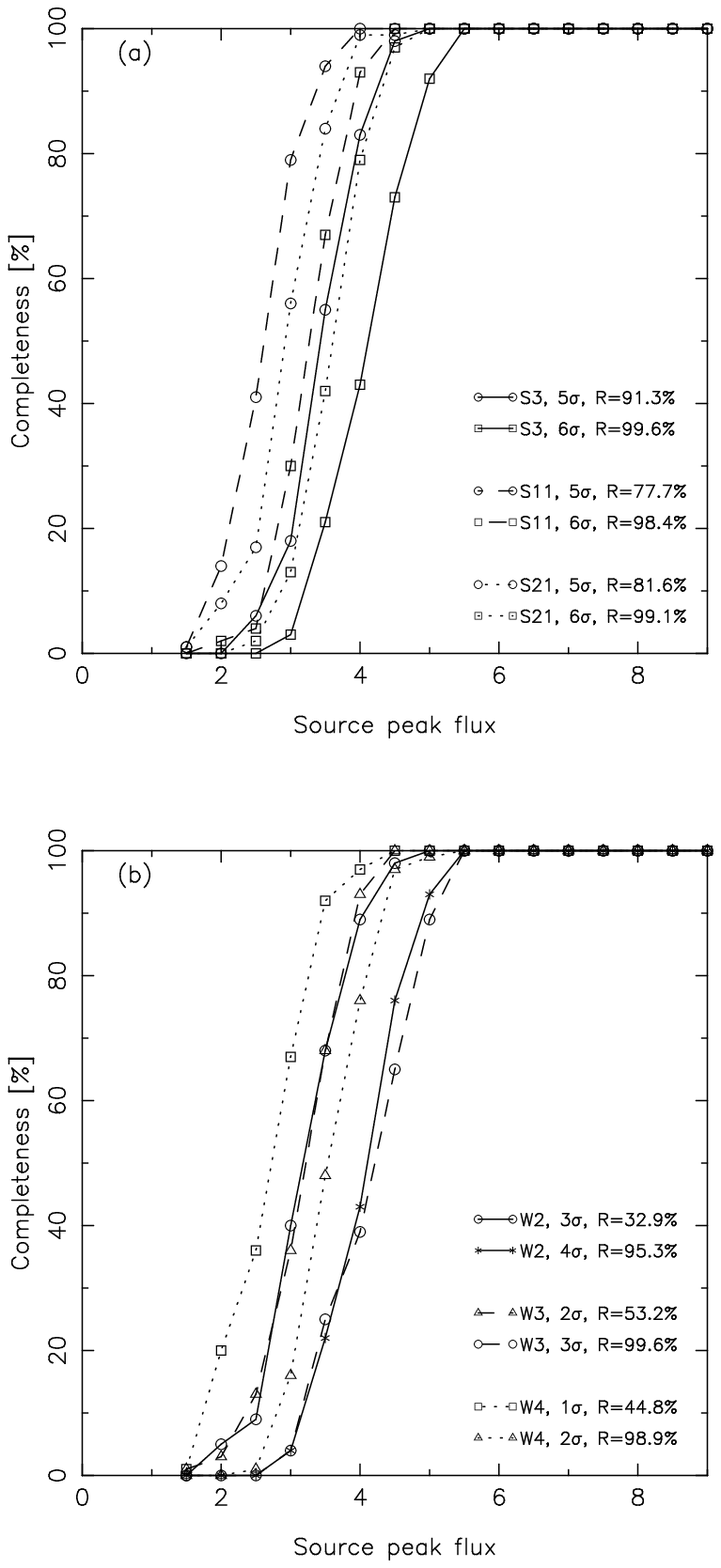}
\caption{As for Fig.~\ref{fig-comp-basic}, but for searches with
different types of pre-processing: (a) smoothing with different
widths: 3, 11 and 21 pixels (b) 1D wavelet reconstruction with
different reconstruction thresholds: $2\sigma$, $3\sigma$ and
$4\sigma$. In each plot, each pre-processing type has its own line
style, while each detection threshold has its own symbol, allowing
comparison of the same detection threshold used with different
pre-processing.}
\label{fig-comp-preproc}
\end{figure}

In Fig.~\ref{fig-comp-basic}, we compare different searches that use
no pre-processing, but only differing degrees of rejection by applying
minimum numbers of channels and spatial pixels. It is apparent that
applying even a small degree of post-finding rejection (rejecting
sources that span only a single channel), the reliability greatly
improves, from 4\% for a $4\sigma$ search to 95\%. The completeness,
however, does reduce: the completeness level for the same
signal-to-noise threshold increases by about 1 unit. This is due to
single-channel matches that align with the true sources - many of
these are probably chance coincidences of noise peaks (which are more
likely to be single-channel sources), and so applying this rejection
will have improved the quality of the resultant
catalogue.

Furthermore, the decrease in the number of spurious sources
means that deeper searches can be done. The deepest one can go with
less than 1 source in 10 spurious is a $6\sigma$ search when accepting
all sources, but by rejecting single-channel sources one can use a
$4\sigma$ search and get at least 1 flux unit deeper for the same
completeness.

Fig.~\ref{fig-comp-preproc} compares a few different types of
preprocessing, together with a basic search rejecting single-channel
sources. Both spectral smoothing and one-dimensional wavelet
reconstruction are considered, each with a range of the critical
parameter (smoothing width or reconstruction threshold).

Fig.~\ref{fig-comp-preproc}(a) shows the spectral smoothing results,
using widths of 3, 11 and 21 channels. These widths span scales below
and above the spectral widths of the sources (Gaussians with FWHM of 5
pixels). The 11-pixel case does the best in terms of completeness, as
it is most closely matched to the shape of the sources.

Fig.~\ref{fig-comp-preproc}(b) shows the results from wavelet
reconstruction case. These are slightly more complicated -- as was seen in
Sec.~\ref{sec-reconExample}, increasing the reconstruction threshold
allows you to search to deeper signal-to-noise thresholds, and
increases the reliability at the expense of completeness for a search
at a given threshold. The symbols in Fig.~\ref{fig-comp-preproc}(b)
relate to the detection threshold, while the lines relate to the
reconstruction threshold.

For the particular simulation under consideration, where the sources have a
single spectral size, the smoothing approach does slightly better
than the wavelet, although the difference in the completeness level is
only about half a flux unit. These tests indicate that getting
complete below about $4\sigma$ is very hard, even with ideally behaved
noise.

\subsection{Processing time}

The other consideration in choosing a processing method is the time
required. This is particularly the case in a survey, where a large
number of cubes may have to be searched, or in the case where a
number of different search types are required. While the utility of
providing execution times is debatable, given the range of
capabilities that are available to researchers, being able to compare
execution times for the same type of search with different processing
options can at least provide some idea of the relative processing
load.

To this end, \duchamp was run using a MacBook Pro (2.66GHz, 8MB RAM)
on the \hipass cube \#201 (of size $170\times160\times1024$ voxels, or
53~MB), with detection thresholds of both $10^8$~Jy~beam$^{-1}$ (no
sources will be found, so that the time taken is dominated by
preprocessing), and 35~mJy~beam$^{-1}$ (or $\sim2.58\sigma$, chosen so
that the time taken will include that required to process sources).
The basic searches, with no pre-processing done, took less than a
second for the high-threshold search, but between 1 and 3~min for the
low-threshold case -- the numbers of sources detected ranged from 3000
(rejecting sources with less than 3 channels and 2 spatial pixels) to
42000 (keeping all sources).

When smoothing, the raw time for the spectral smoothing was only a few
seconds, with a small dependence on the width of the smoothing
filter. And because the number of spurious sources is markedly
decreased (the final catalogues ranged from 17 to 174 sources,
depending on the width of the smoothing), searching with the low
threshold did not add much more than a second to the time. The spatial
smoothing was more computationally intensive, taking about 4 minutes
to complete the high-threshold search.

The wavelet reconstruction time primarily depended on the
dimensionality of the reconstruction, with the 1D taking 20~s, the 2D
taking 30 - 40~s and the 3D taking 2 - 4~min. The spread in times for
a given dimensionality was caused by the different reconstruction
thresholds, with lower thresholds taking longer (since more pixels are
above the threshold and so need to be added to the final spectrum). In
all cases the reconstruction time dominated the total time for the
low-threshold search, since the number of sources found was again
smaller than the basic searches.

\section{Conclusions}

In response to the growing challenges of accurate and reliable
source extraction in large 3D spectral-line datasets, I have developed
\duchamp, a stand-alone source-finding package. Since the noise is the
limiting factor in source extraction, \duchamp uses robust methods to
accurately estimate the noise level, and provides pre-processing
options to prevent the cataloguing of as many noise-generated spurious
sources as possible.

\duchamp has been designed for use with spectral-line surveys in general, and
\hi surveys in particular, but can readily be applied to other types
of data. The core functionality of \duchamp is already forms the basis for
the prototype source finder for the ASKAP processing pipeline, and is
being evaluated, along with other source-finding approaches, by the
ASKAP Survey Science Teams for inclusion in the final pipeline
software. \duchamp is maintained independently, however, and is
provided as a separate, open-source software package.

\section*{Acknowledgements}

Thanks are due to the many people who have provided assistance and
advice during the development and testing of \duchamp, particularly
Ivy Wong, Tobias Westmeier, Mary Putman, Cormac Purcell, Attila
Popping, Tara Murphy, Enno Middelberg, Korinne McDonnell, Philip Lah,
Russell Jurek, Simon Guest, Luca Cortese, David Barnes and Robbie
Auld. Additionally, Emil Lenc and the referee Anita Richards both
provided valuable comments that improved the paper.

\duchamp makes use of the
\textsc{pgplot}\footnote{http://www.astro.caltech.edu/$\sim$tjp/pgplot/},
\textsc{cfitsio}\footnote{http://heasarc.gsfc.nasa.gov/docs/software/fitsio/fitsio.html}
and
\textsc{wcslib}\footnote{http://www.atnf.csiro.au/people/mcalabre/WCS/index.html}
software packages. The bulk of this work was conducted as part of a
CSIRO Emerging Science Postdoctoral Fellowship, and \duchamp continues
to be maintained both as a standalone package and as part of the
software development effort for the ASKAP telescope. This work was
supported by the NCI National Facility at the ANU.

\appendix

\section{How Gaussian noise changes with wavelet scale}
\label{app-scaling}

\begin{table}
\begin{tabular}{llll}
\hline
& $B_3$ Spline & Triangle & Haar\\
& $\{\frac{1}{16},\frac{1}{4},\frac{3}{8},\frac{1}{4},\frac{1}{16}\}$
& $\{\frac{1}{4},\frac{1}{2},\frac{1}{4}\}$
& $\{0,\frac{1}{2},\frac{1}{2}\}$ \\
\hline
\multicolumn{4}{l}{1 dimension}\\
\hline
1     & 0.723489806     & 0.612372436    & 0.707106781   \\
2     & 0.285450405	& 0.330718914    & 0.5           \\
3     & 0.177947535	& 0.211947812    & 0.353553391   \\
4     & 0.122223156	& 0.145740298    & 0.25          \\
5     & 0.0858113122	& 0.102310944    & 0.176776695   \\
6     & 0.0605703043	& 0.0722128185   & 0.125         \\
7     & 0.0428107206	& 0.0510388224   & 0.0883883476  \\
8     & 0.0302684024	& 0.0360857673   & 0.0625        \\
9     & 0.0214024008	& 0.0255157615   & 0.0441941738  \\
10    & 0.0151336781	& 0.0180422389   & 0.03125       \\
11    & 0.0107011079	& 0.0127577667   & 0.0220970869  \\
12    & 0.00756682272	& 0.00902109930  & 0.015625      \\
13    & 0.00535055108	& 0.00637887978  & 0.0110485435  \\
\hline
\multicolumn{4}{l}{2 dimension}\\
\hline
1     & 0.890796310     & 0.800390530     & 0.866025404     \\
2     & 0.200663851	& 0.272878894     & 0.433012702     \\
3     & 0.0855075048	& 0.119779282     & 0.216506351     \\
4     & 0.0412474444	& 0.0577664785    & 0.108253175     \\
5     & 0.0204249666	& 0.0286163283    & 0.0541265877    \\
6     & 0.0101897592	& 0.0142747506    & 0.0270632939    \\
7     & 0.00509204670   & 0.00713319703   & 0.0135316469    \\
8     & 0.00254566946   & 0.00356607618   & 0.00676582347   \\
9     & 0.00127279050   & 0.00178297280   & 0.00338291173   \\
10    & 0.000636389722  & 0.000891478237  & 0.00169145587   \\
11    & 0.000318194170  & 0.000445738098  & 0.000845727933  \\
\hline
\multicolumn{4}{l}{3 dimension}\\
\hline
1     & 0.956543592     & 0.895954449     & 0.935414347 \\
2     & 0.120336499     & 0.192033014     & 0.330718914\\
3     & 0.0349500154    & 0.0576484078    & 0.116926793\\
4     & 0.0118164242    & 0.0194912393    & 0.0413398642\\
5     & 0.00413233507   & 0.00681278387   & 0.0146158492\\
6     & 0.00145703714   & 0.00240175885   & 0.00516748303\\
7     & 0.000514791120  & 0.000848538128 & 0.00182698115\\
\end{tabular}
\caption{Standard deviation scaling coefficients for three different wavelet filter
functions, when used in 1D, 2D and 3D situations. The coefficients
defining each filter are shown at the top of each column.}
\label{tab-scaling}
\end{table}

As discussed in Section~\ref{sec-atrous}, the \atrous algorithm
requires that the thresholds applied at each scale are
equivalent. The threshold is defined as a multiple of the standard
deviation of the original spectrum, scaled appropriately for each
wavelet scale.

However, since the wavelet arrays are produced by convolving the input
array by an increasingly large filter, the pixels in the coefficient
arrays become increasingly correlated as the scale of the filter
increases. This results in the measured standard deviation from a
given coefficient array decreasing with increasing scale. To calculate
this, we need to take into account how many other pixels each pixel in
the convolved array depends on.

To demonstrate, suppose we have a 1-D array with $N$ pixel values
given by $F_i,\ i=1,...,N$, and we convolve it with the B$_3$-spline
filter, defined by the set of coefficients
$\{1/16,1/4,3/8,1/4,1/16\}$. The flux of the $i$th pixel in the
convolved array will be
\[
F'_i = \frac{1}{16}F_{i-2} + \frac{1}{4}F_{i-1} + \frac{3}{8}F_{i}
+ \frac{1}{4}F_{i+1} + \frac{1}{16}F_{i+2}
\]
and the flux of the corresponding pixel in the wavelet array will be
\[
W'_i = F_i - F'_i = \frac{-1}{16}F_{i-2} - \frac{1}{4}F_{i-1}
+ \frac{5}{8}F_{i} - \frac{1}{4}F_{i+1} - \frac{1}{16}F_{i+2}
\]
Now, assuming each pixel has the same standard deviation
$\sigma_i=\sigma$, we can work out the standard deviation for the
wavelet array:
\begin{eqnarray*}
\sigma'_i &= &\sigma \sqrt{\left(\frac{1}{16}\right)^2
+ \left(\frac{1}{4}\right)^2 + \left(\frac{5}{8}\right)^2
+ \left(\frac{1}{4}\right)^2 + \left(\frac{1}{16}\right)^2}\\
&= &0.72349\ \sigma
\end{eqnarray*}
Thus, the first scale wavelet coefficient array will have a standard
deviation of 72.3\% of the input array. This procedure can be followed
to calculate the necessary values for all scales, dimensions and
filters used by \duchamp.

Calculating these values is clearly a critical step in performing the
reconstruction. The method used by \citet{starck02a} was to simulate
data sets with Gaussian noise, take the wavelet transform, and measure
the value of $\sigma$ for each scale. We take a different approach, by
calculating the scaling factors directly from the filter coefficients
by taking the wavelet transform of an array made up of a 1 in the
central pixel and 0s everywhere else. The scaling value is then
derived by taking the square root of the sum (in quadrature) of all
the wavelet coefficient values at each scale. We give the scaling
factors for the three filters available to \duchamp in
Table~\ref{tab-scaling}. These values are hard-coded into \duchamp, so
no on-the-fly calculation of them is necessary.

Memory limitations prevent us from calculating factors for large
scales, particularly for the three-dimensional case (hence the smaller
table). To calculate factors for higher scales than those available,
we divide the previous scale's factor by either $\sqrt{2}$, $2$, or
$\sqrt{8}$ for 1D, 2D and 3D respectively.

\section{Example of an output catalogue}
\label{app-example}

Fig.~\ref{fig-exampleCat} shows an example of the ASCII output table
produced by \duchamp. This table comes from the HIPASS cube \#201,
processed with 1D wavelet reconstruction (with a $3\sigma$
reconstruction threshold) and searched to a $4\sigma$ threshold (to
provide a small catalogue for demonstration purposes).

\begin{figure*}
\begin{minipage}{170mm}
{\small
\begin{verbatim}
-------------------------------------------------------------------------------------------
 Obj#           Name     X     Y     Z           RA          DEC      VEL     w_RA    w_DEC
                                                                   [km/s] [arcmin] [arcmin]
-------------------------------------------------------------------------------------------
    1 J060921-215712  59.4 140.5 114.6  06:09:21.89 -21:57:12.54  225.613    64.57    43.20
    2 J060229-254657  83.4  83.3 118.3  06:02:29.09 -25:46:57.48  274.754    40.05    27.94
    3 J060604-272027  71.4  59.8 121.3  06:06:04.62 -27:20:27.38  314.287    72.41    51.54
    4 J061118-213700  52.5 145.4 162.5  06:11:18.70 -21:37:00.27  860.075    32.39    23.49
    5 J060033-285911  89.7  35.2 202.1  06:00:33.77 -28:59:11.64 1386.318    27.91    28.12
    6 J061706-272350  34.7  58.3 227.8  06:17:06.48 -27:23:50.26 1728.877    24.89    27.30
    7 J055848-252527  95.8  88.6 231.8  05:58:48.61 -25:25:27.83 1782.146    31.86    24.19
    8 J061522-263426  40.2  70.8 232.6  06:15:22.77 -26:34:26.29 1792.992    16.53    19.59
    9 J060053-214235  88.9 144.3 232.7  06:00:53.49 -21:42:35.96 1794.159    35.95    28.16
   10 J060443-260638  75.8  78.3 233.4  06:04:43.20 -26:06:38.58 1803.162    24.13    23.87
   11 J060107-234004  88.0 115.0 235.5  06:01:07.33 -23:40:04.33 1831.452    35.95    36.09
   12 J061213-214901  49.4 142.3 271.3  06:12:13.72 -21:49:01.21 2309.807    28.48    27.52
   13 J061617-213333  35.2 145.9 298.4  06:16:17.30 -21:33:33.92 2673.464    20.21     7.47
   14 J060838-223918  62.0 130.0 724.9  06:08:38.82 -22:39:18.06 8510.966     4.05     3.96
   15 J055212-291706 117.0  30.5 727.0  05:52:12.72 -29:17:06.49 8539.914    15.50    24.34

-------------------------------------------------------------------------------------------
    w_50    w_20   w_VEL     F_int     F_tot    F_peak S/Nmax  X1  X2  Y1  Y2  Z1  Z2  Npix
  [km/s]  [km/s]  [km/s] [Jy km/s] [Jy/beam] [Jy/beam]                                [pix]
-------------------------------------------------------------------------------------------
  26.003  64.416  52.851    13.932    15.486     0.213  19.65  52  67 135 145 114 118   211
  24.620  59.902  66.084     3.544     3.938     0.118  10.90  78  87  80  86 117 122    67
  31.622  46.235 105.758    13.622    15.133     0.150  13.83  65  82  53  65 118 126   248
  87.170 110.181 106.141    30.126    33.345     0.410  37.94  49  56 142 147 159 167   266
 178.136 196.548 173.083    17.993    19.846     0.173  15.98  87  93  32  38 196 209   295
 307.095 327.361 293.519     7.198     7.921     0.093   8.58  32  37  55  61 215 237   146
 237.471 274.205 240.237    10.753    11.830     0.115  10.67  92  99  86  91 221 239   215
  70.682  90.968  53.400     1.671     1.838     0.068   6.32  39  42  69  73 231 235    35
 103.364 227.932 226.900    21.574    23.731     0.166  15.30  85  93 141 147 222 239   336
 215.186 232.314 213.600    17.760    19.536     0.155  14.28  73  78  76  81 225 241   305
 201.838 254.227 267.072    60.536    66.574     0.297  27.44  84  92 111 119 226 246   742
 409.395 530.057 549.514    13.523    14.825     0.101   9.31  46  52 140 146 257 298   270
 142.639 249.637 214.911     2.879     3.148     0.127  11.74  33  37 145 146 294 310    41
 141.561 249.361 125.554     0.497     0.523     0.081   7.51  62  62 130 130 720 729     8
 221.143 304.807 362.794    26.690    28.089     0.479  44.26 116 119  28  33 714 740   135

-----------------------------------------------------------------
 Flag  X_av  Y_av  Z_av X_cent Y_cent Z_cent X_peak Y_peak Z_peak
                                                                 
-----------------------------------------------------------------
    -  59.7 140.5 114.7   59.4  140.5  114.6     59    140    114
    -  83.4  83.2 118.3   83.4   83.3  118.3     84     83    118
    -  71.5  59.8 121.3   71.4   59.8  121.3     72     61    121
    E  52.5 145.1 162.5   52.5  145.4  162.5     53    146    164
    -  89.7  35.2 202.2   89.7   35.2  202.1     90     36    197
    -  34.7  58.3 227.8   34.7   58.3  227.8     35     58    236
    -  95.8  88.6 231.5   95.8   88.6  231.8     96     89    237
    -  40.1  70.8 232.6   40.2   70.8  232.6     40     71    231
    E  88.8 144.2 232.3   88.9  144.3  232.7     89    144    233
    -  75.9  78.3 233.4   75.8   78.3  233.4     76     78    240
    -  88.0 114.9 235.7   88.0  115.0  235.5     88    115    231
    -  49.4 142.3 271.8   49.4  142.3  271.3     50    142    259
    E  35.3 145.8 298.7   35.2  145.9  298.4     34    146    297
    -  62.0 130.0 724.8   62.0  130.0  724.9     62    130    723
    - 117.0  30.3 726.7  117.0   30.5  727.0    117     30    733

\end{verbatim}
}
\caption{Example of an ASCII-format output catalogue from
  \duchamp. Note that the columns starting with X, Y, and Z are all in
  pixel units. The file has been split into three sections to enable
  it to fit on the page.}
\label{fig-exampleCat}
\end{minipage}
\end{figure*}

\label{lastpage}

\end{document}